\documentclass[11pt]{article}

\newcommand{\ifabs}[2]{#2}
\renewcommand{\ifabs}[2]{#1}

\ifabs{}{}

\usepackage{comment}
\usepackage{cite}
\usepackage{fullpage}
\usepackage{color}
\usepackage{epic}
\usepackage{eepic}
\usepackage{epsfig}
\usepackage{xspace}
\usepackage{graphicx}

\ifabs{
\excludecomment{fullproof}
\newenvironment{proofsketch}{\noindent\par{\bf Sketch of Proof: }}{\nopagebreak\rule{1 ex}{0.8 em}\medskip}
\includecomment{abs}
\excludecomment{fullpaper}

}{

\excludecomment{proofsketch}
\excludecomment{abs}
\includecomment{fullpaper}
}

\excludecomment{suppress}

\newcommand{\todo}[1]{\typeout{TODO: \the\inputlineno: #1}\textbf{[[[ #1 ]]]}}
\newcommand{\concept}[1]{\textbf{#1}}

\newtheorem{theorem}{Theorem}
\newtheorem{lemma}[theorem]{Lemma}
\newtheorem{corollary}[theorem]{Corollary}

\newcommand{\newloglike}[2]{\newcommand{#1}{\mathop{\rm #2}\nolimits}}
\newloglike{\E}{E}
\newloglike{\sgn}{sgn}

\newcommand{\etal}[1]{{\it et al.\/}}

\newenvironment{proof}{\noindent\par{\bf Proof: }}{\nopagebreak\rule{1 ex}{0.8 em}\medskip}

\newcommand{\floor}[1]{\left\lfloor{#1}\right\rfloor}

\newcommand{\makespan}{C^{\max}}
\newcommand{\reassignment}{r}

\newcommand{\binhash}{\emph{BinHash}\xspace}

\newcommand{\OPT}{\mbox{OPT}}

\begin{document}

\title{Path-Independent Load Balancing With Unreliable Machines}

\author{
James Aspnes\thanks{Department of Computer Science,
Yale University.}~\thanks{Supported in part by NSF
  grants CNS-0305258 and CNS-0435201.  Email:
  \texttt{aspnes@cs.yale.edu}.}
\and Yang Richard Yang\footnotemark[1]~\thanks{Supported in part by NSF
  grants CNS-0435201 and ANI-0238038.  Email:
  \texttt{yry@cs.yale.edu}.}
\and Yitong Yin\footnotemark[1]~\thanks{Supported by NSF grant
CNS-0305258.  Email: \texttt{yitong.yin@yale.edu}.}
}

\maketitle

\begin{abstract}
We consider algorithms for load balancing on unreliable machines.
The objective is to optimize the two criteria of minimizing the makespan
 and minimizing job reassignments in response to machine failures.
We assume that the set of jobs is known in advance but that the
pattern of machine failures is unpredictable.  Motivated by the
requirements of BGP routing, we consider
\concept{path-independent} algorithms, with the property that the job
assignment is completely determined by the subset of available
machines and not the previous history of the assignments.
We examine first the question of performance measurement of path-independent
load-balancing algorithms, giving the measure of makespan and the normalized measure of 
reassignments cost.  We then describe two classes of algorithms for
optimizing these measures against an oblivious adversary for
identical machines.
The first, based on independent random
assignments, gives expected reassignment costs within a factor of $2$
of optimal
and gives a makespan within a factor of
$O(\log m/\log \log m)$ of optimal with high probability, for
unknown job sizes.
The second, in which jobs are first grouped into bins and at most one bin
is assigned to each machine,
gives constant-factor ratios on both reassignment cost and makespan, for
known job sizes.
Several open problems are discussed.
\end{abstract}

\ifabs{
\setcounter{page}{0}
\thispagestyle{empty}
\vfill
\pagebreak
}

\section{Introduction}
\label{section-introduction}

Given a set of jobs $J = \{1\ldots n\}$ and machines $M = \{1 \ldots
m\}$, where each
job $j$ has a \concept{size} or \concept{processing time} $p_{j}$,
the problem of \concept{load
balancing} is to construct an assignment $f: J \rightarrow M$ that
minimizes the \concept{makespan}, defined as $\makespan = \max_{i \in M} \sum_{j
\in f^{-1}(i)} p_{j}$.\footnote{In the classic $\alpha|\beta|\gamma$
notation of Graham~\etal~\cite{GrahamLLK1979}, we are considering
primarily the $P||\makespan$ scheduling problem.  Our use of notation
largely follows that of~\cite{Leung2004}.}

We consider a variant of load balancing in which the set of
\concept{available} machines $S$ changes over time.  In effect, we are
solving a sequence of $P||\makespan$ scheduling problems where the set
of jobs is fixed but the set of machines varies, and our goal is
to minimize both the makespan $\makespan$ and the \concept{reassignment
cost} of moving jobs from one machine to another as new machines
become available and old machines leave.  As a further complication,
we restrict ourselves to
\concept{path-independent} algorithms, those that always assign the
same jobs to the same machines given a particular set $S$ despite any
previous history of assignments.  This restriction simplifies the
description of an algorithm, since we can just present an assignment
for each nonempty set of machines, but it may dramatically increase
reassignment costs since we cannot take previous assignments into
account.  Surprisingly, we show that randomization can lower the expected
reassignment cost between any two states of a path-independent algorithm to within a
constant factor of optimal, while maintaining a constant
approximation ratio to the optimal makespan.

This variant of load-balancing is
inspired by the problem of routing among multiple network paths using
the Border Gateway Protocol (BGP)~\cite{BGP}, the {\it de facto}
interdomain routing protocol of the Internet. In BGP, the global
Internet is divided into multiple autonomous systems (AS). Each AS has
several peering ASes, and the ASes export to its peers available routes
to destination prefixes (jobs). Each AS maintains a cache (called
routing information base) of currently available routes to the
destinations exported by its peers, and selects routes to destinations
from the routing cache using a route selection algorithm. One major
objective of the route selection algorithm is load balancing
among multiple peering links
(machines)~\cite{WXYLe05}. Figure~\ref{fig:bgp-model} shows the standard
protocol/process model of interdomain route selection of
BGP~\cite{CR05,Sob03,WXYLe05}.
\begin{figure}[htb]
%\centerline{\psfig{figure=egress-model.eps,width=2.5in,angle=0}}
\centerline{\includegraphics[width=2.5in]{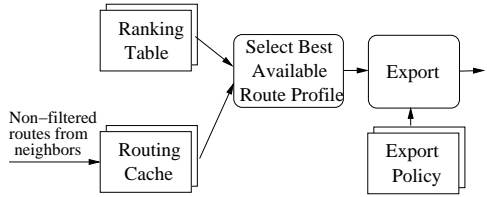}}
\caption{The protocol/process model of route selection for interdomain
traffic engineering of BGP.}
\label{fig:bgp-model}
\end{figure}

Because network connections are dynamic, the set of routes that are
available at any time may vary. When the set of available routes
changes, BGP will re-run the route selection algorithm. If the
assignment of the route to a destination is changed, BGP will update its
peers, and this update can propagate throughout the global
Internet. Thus, it is important to minimize the number of reassignments,
since they are expensive operations both in terms of bandwidth and
router CPU processing~\cite{SCEH05}.

Naturally, some reassignments are unavoidable: when a
peering link goes down, all destination prefixes that previously used that link
must be assigned elsewhere, and when new links become
available, we cannot refuse to reassign destinations to them without
violating our load-balancing criterion.  So the number of
reassignments performed by an
algorithm will need to be measured against an estimate of how many
reassignments are necessary, which will be obtained by considering
the behavior of an ideal algorithm that maintains perfect load
balance with minimal reassignment costs.

A further complication is that routers (e.g., cisco and
Juniper routers) implement route selection by assigning priorities
(called local preference values) to routes. This means that the
assignments of destinations  to peering links are unaffected by
past history, a property known as
\concept{path-independence}. Thus, we will consider
only load-balancing algorithms in which the current assignment is
determined only by the current set of available
machines.  Furthermore,
for multihomed stub ASes, such path-independent assignments can
guarantee interdomain routing \concept{stability}~\cite{WXYLe05}.
Path-independence is thus a desirable property of load-balancing
algorithms for unreliable machines, motivating our study of such algorithms.

Formally, a path-independent load-balancing algorithm is just a
function from $2^M
\setminus
\{ \emptyset \}$ to $M^J$, where the argument represents the (always
nonempty) set of available machines and the function value represents an
assignment, with the constraint that no job is ever assigned to a
machine that is not in the available set.  Given such an algorithm $A$,
we write $A_S$ for the assignment chosen by $A$ for available set $S$
and $A_S(j)$ for the machine to which job $j$ is assigned.  With this
notation, the constraint on $A$ is that $A_S(j) \in S$ for all $S$ and
$j$.

Because the assignment of jobs to machines is determined by the set
of available machines, this set completely determines the state of
the system.  We thus identify sets of available machines with states
of the system and will refer to such sets simply as states hereafter.

\subsection{Our results}
\label{section-results}

The paper contains several novel contributions:

\begin{itemize}
\item The new problem of \concept{path-independent load balancing},
motivated by practical issues in BGP routing.
\item A measure of \concept{reassignment costs} that takes into
account the difference between small and large changes in the set of
available machines.  This measure is motivated and described in
Section~\ref{section-measuring-performance}.  We also show that
standard competitive analysis techniques~\cite{SleatorT1985} are
not useful.
\item A simple algorithm, based on independent random machine
preferences, that achieves low reassignment costs and $O(\log m/\log
\log m) \cdot \OPT$ makespan even with unknown job sizes.  This
algorithm, together with a more general class of
\concept{preference-based algorithms} of which it is a member, is described in
Section~\ref{section-preference-based-algorithms}.
\item A more sophisticated algorithm \binhash for jobs
of known sizes, based on consolidating jobs into
a variable number of bins (depending on the number of available
machines) and then hashing the bins.  This algorithm achieves constant approximation
ratios on both reassignment costs and makespan.  It is
described in Section~\ref{section-binhash}.
\end{itemize}

Finally, in Section~\ref{section-conclusions} we discuss
possibilities for future work.

\ifabs{Because of space limitations, some proofs are sketched or moved
to the appendix.  Complete proofs can be found in the full paper.}{}

\subsection{Related work}
\label{section-related-work}

It has long been known that a simple greedy
algorithm~\cite{Graham1966} achieves a makespan within a factor of $2$
of optimal on identical machines.  Much recent work on the problem has
focused on on-line load-balancing, where jobs arrive one at a time;
see~\cite{Albers2002} for an example of the current state of the art.
Our work is distinguished from this work by the assumption that jobs
are known, but that the set of available machines changes over time.

Kalyanasundaram and Pruhs~\cite{KalyanasundaramP2000,KalyanasundaramP2005}
have
considered models of fault-tolerant scheduling for parallel computers;
here the issue is that process failure prevents any job assigned to it
from being completed and the goal is to maximize the total value of
completed jobs subject to release time and completion time
constraints.  Their results are based on redundant scheduling without
pre-emption and are not directly applicable to our problem.

The technique of consolidating jobs in our algorithm for
known job sizes is similar to an approach taken by
Sibeyn~\cite{Sibeyn1991} for load-balancing jobs with sizes drawn
from a random distribution.  Sibeyn's techniques are intended to
reduce the variance in job sizes and do not have the
low-reassignment-cost properties of our prefix-based method.

There has been substantial work on load-balancing mechanisms based on
the \concept{power of two choices}, where jobs pick two (or more)
machines at random and choose the more lightly-loaded machine.
(See~\cite{MitzenmacherRS2000} for a survey of these results.)  We
found through experiments that this approach, while producing excellent
makespans, does not appear to yield low reassignment costs: the chains
of displaced jobs that migrate to less loaded machines when a new
machine becomes available are simply too long.

\section{Measuring performance}
\label{section-measuring-performance}

We measure the performance of a path-independent load-balancing algorithm by
two criteria: the \concept{makespan} of $A_S$ for each set of
available machines $S$, and the \concept{reassignment cost} paid by
$A$ when moving from one assignment $A_S$ to a different assignment
$A_{T}$.

The makespan was is defined using standard notation as
$\makespan_A(S) = \max_{i \in S} \sum_{j \in A_S^{-1}(i)} p_{j}$, the
maximum total load on any one machine.

The reassignment cost $\reassignment_A(S,T)$
is the number of jobs that
move from one machine to another between $A_S$ and $A_{T}$.
Formally, we define
$\reassignment_A(S,T) = |\{ j | A_S(j) \neq A_{T}(j)\}|$.
Note that sizes are not used in computing the reassignment cost; we
assume that all jobs incur an equal overhead when reassigned
regardless of size.  Note also that reassignment cost is
symmetric: $\reassignment_A(S,T) = \reassignment_A(T,S)$ for all $S$ and $T$.

An \concept{execution} of a path-independent load-balancing algorithm is specified as
a sequence of states $S_0, S_1, S_2, \ldots$, which may or may not be
finite.  For a finite execution $S_0 \ldots S_t$,
the \concept{total reassignment cost} of an algorithm $A$ is
$\sum_{i=0}^{t-1} \reassignment_A(S_i, S_{i+1})$.
For an infinite execution, the \concept{average reassignment cost} is
defined as
$\lim_{t \rightarrow \infty}
 \frac{1}{t} \sum_{i=0}^{t-1} \reassignment_A(s_i,s_{i+1})$.
Note that because $\reassignment_A$ is bounded by $n$ this limit always exists.

Because both makespan and reassignment cost are parameterized by
states, an algorithm's overall performance may depend strongly on the
details of a specific execution.
To be able to compare algorithms it
will be useful to have summary statistics that describe the
performance of an algorithm over all possible executions.
Worst-case measures are of limited usefulness: the worst-case
reassignment cost of any algorithm on two or more machines is $n$
jobs per step, since we may have only a single available machine at
each step that alternates between steps, forcing us to always
reassign jobs.  Similarly, the worst-case makespan which arises when
there is only one available machine does not distinguish algorithms
in any way.  So instead, we must consider measures that take into
account the difficulty of particular executions.

We will show first that a traditional competitive analysis approach
does not suffice for this purpose, and then propose an alternative
method where the cost of an execution is normalized by a straw-man
\concept{reference cost} corresponding to the performance of a
particular idealized load-balancing algorithm.

\subsection{Impossibility of bicriteria-competitive algorithms}
\label{section-impossibility-of-bicriteria-competitive-algorithms}

The technique of \concept{competitive analysis}~\cite{SleatorT1985}
compares the cost of a candidate on-line algorithm (which receives the
input only step-by-step) in a given history against the cost of an
optimal off-line algorithm (which sees the entire input in advance and
is typically not computationally bounded).  A candidate algorithm is
said to have \concept{competitive ratio} $c$ or be $c$-competitive if
its cost is at most $c$ times the cost of the optimal algorithm, plus
a constant.\footnote{The constant allows excluding perverse effects of
very short executions.}

For the path-independent load-balancing problem, we are measuring two
quantities at each step: the makespan in the new state, and the
reassignment cost between the old state and the new state.  A natural
way to apply competitive analysis to this situation might be to
adopt a bicriteria approach and insist that our candidate algorithm
be $(\alpha,\beta)$-competitive, where $\alpha$ bounds the ratio
between the candidate's makespan in each state and the optimal algorithm's
makespan in its corresponding state,
and $\beta$ bounds the ratio between the candidate's total
reassignment cost over a finite execution and the optimal algorithm's total
reassignment cost.\footnote{The reason for using a worst-state ratio
for makespan but a total ratio for reassignment costs is that the
algorithm could spend an arbitrarily long time in a bad state, while
reassignments costs accumulate more straightforwardly over
transitions.  We could also demand that the candidate's
reassignment cost is at most $\beta$ times the optimal algorithm's
reassignment for each transition, but this only makes life harder for
the candidate algorithm and excludes algorithms that depend on
amortizing reassignment costs.}  We show below that no such algorithm
exists for any finite $\alpha$ and $\beta$ in any system with
$n \ge 2$ jobs and $m \ge 2$ machines,
even if the jobs have identical sizes.

\begin{theorem}
\label{theorem-no-competitive-algorithm}
Let $A$ be a path-independent load-balancing algorithm for a system with
$n \ge 2$ jobs and $m \ge 2$ machines.
Then $A$ is not $(\alpha,\beta)$-competitive for
any $\alpha$ and $\beta$.
\end{theorem}
\begin{proof}
Without loss of generality, let $m=2$ (we can always make any
additional machines permanently unavailable).

Let $A$ be some candidate algorithm.  We will show that $A$ is not
competitive with respect to reassignment costs regardless of
makespan.  Let $i \in \{1,2\}$ be such that some job is assigned to
machine $i$ in $A_{\{1,2\}}$; let $i'$ be the other machine. Now
consider an execution in which $S_t = \{1,2\}$ when $t$ is even and
$S_t = \{i'\}$ when $t$ is odd. Because there is some job that is
assigned to machine $i$ at even times and not odd times, $A$ pays a
reassignment cost of at least $1$ per step of the execution.
However, an optimal $A^*$ that assigns all jobs to machine $i'$ in
all states pays $0$.  No finite $\beta$ nor additive constant is
large enough to overcome in the limit the infinite ratio between $A$
and $A^*$'s reassignment costs.
\end{proof}

\subsection{Normalized reassignment costs}
\label{section-normalized-costs}

We define the \concept{normalized reassignment cost}
relative to the costs of an ideal algorithm that always maintains
perfect balance.  We consider first the case of $n=km!$ identical jobs,
where $k$ is some integer; the $m!$ factor ensures that the jobs
can be equally divided over any subset of the machines.
In this case we can directly calculate the reassignment cost when
moving from state $S$ to state $T$.

\begin{theorem}
\label{theorem-normalized-reassignment}
Consider a system $m$ machines and $n=km!$ identical jobs.  Let
$A$ be an algorithm that assigns exactly $n/|S|$ jobs to each
machine in each state $S$.  Then the number of reassignments performed by $A$
going from $S$ to $T$ is at least
\begin{equation}
\label{eq-normalized-reassignment}
\reassignment^*(S,T) =
n \cdot
\left(
    1 - \frac{|S \cap T|}{\max\left(|S|, |T|\right)}
\right).
\end{equation}
\end{theorem}
\begin{proof}
Consider the set of jobs assigned to machines in $S \setminus T$ by
$A_S$; since $A$ assigns exactly $n/|S|$ jobs to each machine in $S
\setminus T$,
there are $n\cdot|S \setminus T|/|S|$ such jobs total.  All of
these jobs must be reassigned going from $S$ to $T$.

Conversely, any job assigned to a machine in $T \setminus S$ by
$A_T$ must also have been reassigned going from $S$ to $T$.  Though
these two sets of jobs may overlap, $A$ must reassign at least
\begin{eqnarray*}
n \cdot
\max
\left(
    \frac{|S \setminus T|}{|S|},
    \frac{|T \setminus S|}{|T|}
\right)
&=&
n \cdot
\max
\left(
    \frac{|S \setminus (S \cap T)|}{|S|},
    \frac{|T \setminus (S \cap T)|}{|T|}
\right)
\\
&=&
n \cdot
\max
\left(
    1 - \frac{|S \cap T|}{|S|},
    1 - \frac{|S \cap T|}{|T|}
\right)
\\
&=&
n \cdot
\left(
    1 - \frac{|S \cap T|}{\max\left(|S|, |T|\right)}
\right).
\end{eqnarray*}
\end{proof}

We will take $\reassignment^*$ as the ideal reassignment cost, and measure the
reassignment cost of any particular algorithm $A$ as the maximum over all
$S$ and $T$ of the ratio of its reassignment cost $\reassignment_A(S,T)$ to
$\reassignment^*(S,T)$.

A justification for this approach is that
the $\reassignment^*$ lower bound continues to hold in expectation for \emph{any}
algorithm---regardless of whether it distributes jobs evenly or
not---provided
the machine names are randomly permuted before the
algorithm is used and such renaming is undone afterwards.
This fact is
formally stated in the following theorem.
The random permutation of machine names, which is easily implemented
by an oblivious adversary, prevents an algorithm from
getting lucky by placing all of its jobs on machines that stay
available in both $S$ and $T$.

\begin{theorem}
\label{theorem-randomized-lower-bound} For any algorithm $A$ that
maps states to assignments, choose a permutation $\rho$ uniformly at
random. Let $\rho^{-1} A\rho$ be the algorithm that constructs an
assignment for $S$ by applying $A$ to $\rho S$ and then undoing
the machine renaming. Then the algorithm $\rho^{-1} A\rho$ reassigns
at least \mbox{$n\cdot\left(1- \frac{|S \cap
T|}{\max(|S|, |T|)}\right)$} jobs in expectation when moving from state $S$ to
state $T$.
\end{theorem}
\newcommand{\TheoremRandomizedLowerBoundProof}{
Fix some particular job $j$. Let $A_S(j)$ be the machine to which $A$
assigns job $j$ in state $S$. Note that $A_S(j)\in S$.
Observe that:
\begin{itemize}
\item For any $\langle U,V\rangle$ where $|U|=|S|$, $|V|=|T|$ and
  $|U\cap V|=|S\cap T|$, there exists some permutation
  $\rho$ such that $\langle U,V\rangle=\langle \rho S,\rho T\rangle$.

\item There are ${m\choose |S|}{|S|\choose |S\cap T|}{m-|S|\choose
  |T|-|S\cap T|}$ equivalence classes after applying permutations to
  $\langle S,T\rangle$, each of which contains $|S\cap T|!(|S|-|S\cap
  T|)!(|T|-|S\cap T|)!(m-|S\cup T|)!$ permutations. We denote this
  value as $e(S,T)$.

\item Fix $S$ and $T$. For any $\rho S$, the number of $\rho T$ which
  contain $A_{\rho S}(j)$, i.e., $|\{\rho T\mid A_{\rho S}(j)\in\rho
  T\}|$ equals ${|S|-1\choose |S\cap T|-1}{m-|S|\choose |T|-|S\cap T|}$.
\end{itemize}

Then the probability that job $j$ stays while the state going from $S$ to
$T$
\begin{eqnarray*}
  \Pr\rho[A_{\rho S}(j)=A_{\rho T}(j)] & = &
  \frac{1}{m!}\sum_\rho|\{\rho\mid A_{\rho S}(j)=A_{\rho T}(j)\}|\\
  & = & \frac{1}{m!}\sum_{\rho S}|\{\rho T\mid A_{\rho S}(j)=A_{\rho T}(j)\}|\cdot e(S,T)\\
  & \le & \frac{1}{m!}\sum_{\rho S}|\{\rho T\mid A_{\rho S}(j)\in\rho
  T)\}|\cdot e(S,T)\\
  & = & \frac{1}{m!}\sum_{\rho S}{|S|-1\choose |S\cap
  T|-1}{m-|S|\choose |T|-|S\cap T|}\cdot e(S,T)\\
  & = & \frac{|S\cap T|}{|S|}.
\end{eqnarray*}
Symmetrically,
\begin{eqnarray*}
  \Pr\rho[A_{\rho S}(j)=A_{\rho T}(j)] & \le & \frac{|S\cap T|}{|T|}.
\end{eqnarray*}
Therefore, we have $\Pr\rho[A_{\rho S}(j)=A_{\rho T}(j)]\le\frac{|S\cap
  T|}{\max(|S|,|T|)}$. The probability that job $j$ is reassigned
  from $S$ to $T$ is thus lower bounded by $(1-\frac{|S\cap
  T|}{\max(|S|,|T|)})$. The expected total cost is obtained by summing
  over all $n$ jobs.
}
\ifabs{

The proof is given in
Appendix~\ref{appendix-randomized-lower-bound-proof}.
}{\begin{proof}
\TheoremRandomizedLowerBoundProof
\end{proof}}

\section{Preference-based algorithms}
\label{section-preference-based-algorithms}

A \concept{preference-based algorithm} is one in which each job is
assigned a permutation of the machines (which may depend on the set
of other jobs), and always moves to the first available machine in its
permutation.  We can think of the operation of a preference-based
algorithm as choosing for each job $j$ and state $S$ the machine $i$
in $S$ that minimizes $\sigma_j(i)$, where $\sigma_j$ is the
permutation for job $j$.

We consider first the reassignment costs of preference-based
algorithms in general and then the makespan of the simple
preference-based algorithm where each $\sigma_j$ is a random
permutation.

\subsection{Reassignment costs}
\label{section-preference-based-reassignment-costs}

Preference-based algorithms have the desirable property that they
achieve close to the minimum reassignment cost against an oblivious adversary,
provided the preferences are permuted randomly before the algorithm
is used.  This fact is stated formally in the following theorem.

\begin{theorem}
\label{theorem-preference-based-reassignment}
For each job $j$, let $\sigma_j$ be a permutation of the machines.
Choose a permutation $\rho$ uniformly at random.  Then the
preference-based algorithm using preferences $\sigma_j \rho$
reassigns an expected
$n \cdot \left(1 - \frac{|S \cap T|}{|S \cup T|}\right)$ jobs
when moving from state $S$ to state $T$.
\end{theorem}
\begin{proof}
Fix some particular job $j$.  Going from $S$ to $T$, job $j$ stays
put just in case
$\min_{i \in S} \sigma_j \rho(i) = \min_{i \in T} \sigma_j \rho(i)$.
This occurs precisely
when $\min_{i \in S \cup T} \sigma_j \rho(i)$ is achieved by some $i$
in $S \cap T$, i.e. when neither $S$ nor $T$ provides a machine that
$j$ prefers to the best machine in their intersection.
Since $\sigma_j \rho$ is a random permutation, the probability that
$j$ does \emph{not} move is thus precisely $|S \cap T| / |S \cup T|$.
The expected number of moves is obtained by taking one minus this probability
and summing over all $n$ jobs.
\end{proof}

\begin{corollary}
\label{corollary-preference-based-reassignment}
For any preference-based algorithm $A$ and any states $S$ and $T$,
$\E[\reassignment_A(S,T)] \le 2 \reassignment^*(S,T)$, where $\reassignment^*$ is defined as in
Theorem~\ref{theorem-normalized-reassignment}.
\end{corollary}
\ifabs{
\begin{proofsketch}
Follows from showing that 
$
\E[\reassignment_A(S,T)]
=
n \cdot \left(1 - \frac{|S \cap T|}{|S \cup T|}\right)
\le
2n \cdot \max
\left(
    \frac{|S \setminus T|}{|S|},
    \frac{|T \setminus S|}{|T|}
\right)
= 2\reassignment^*(S,T)$.
\end{proofsketch}
}{
\begin{proof}
Observe that
\begin{eqnarray*}
\E[\reassignment_A(S,T)]
&=&
n \cdot \left(1 - \frac{|S \cap T|}{|S \cup T|}\right)
\\
&=&
n \cdot \left(\frac{|S \setminus T| + |T \setminus s|}
{|S \cup T|}\right)
\\
&=&
n \cdot
\left(
    \frac{|S \setminus T|}{|S \cup T|}
+
    \frac{|T \setminus S|}{|S \cup T|}
\right)
\\
&\le&
n \cdot
\left(
    \frac{|S \setminus T|}{|S|}
+
    \frac{|T \setminus S|}{|T|}
\right)
\\
&\le&
2n \cdot \max
\left(
    \frac{|S \setminus T|}{|S|},
    \frac{|T \setminus S|}{|T|}
\right)
\\
&=&
2\reassignment^*(S,T).
\end{eqnarray*}
\end{proof}
}

\subsection{Makespan for random preferences}
\label{section-makespan-for-random-preferences}

A typical case of preference-based algorithms is the random
preference algorithm, in which a fixed random permutation of
machines is picked independently for each job as its preference
list.

This randomization implicitly subsumes the random permutation
of machine names assumed in
Corollary
\ref{corollary-preference-based-reassignment}; thus the corollary
applies and random preferences yield an
expected $2\reassignment^*(S,T)$ reassignments going from state $S$
to $T$.  For \emph{identical} jobs, the makespan of random preference
when all machines are available is
$\Theta(\log m/\log\log m) \cdot \OPT$ \emph{w.h.p.}~for $n=m$, and is
$(n/m)+\Theta\left(\sqrt{(n/m)\log m}\right)
= \OPT \cdot \left(1 + \Theta\left(\sqrt{(m/n) \log m}\right)\right)$
\emph{w.h.p.}~for large $n$, using
standard balls-in-bins results (e.g.~\cite{RaabS1998}).
Reducing the number of available machines effectively replaces $m$
with $|S|$ in the latter expression.

If the jobs are not identical, we can still argue that the makespan
is $O(\log m / \log \log m) \cdot \OPT$, using a generalization
to weighted balls of the usual balls-in-bins results.  Here we assume
that we have a bound on both the weight of the largest ball and the
expected weight of the balls assigned to any one bin.  These
quantities are both bounded by the optimal makespan.

\begin{lemma}
\label{lemma-weighted-balls-in-bins}
Let $n$ balls with non-negative weights $w_1, w_2, \ldots w_n$ be distributed
independently and uniformly at random into $b \le m$ bins.  Let $W =
\max\left(\max(w_i), \frac{1}{b} \sum_{i=1}^{n} w_i\right).$
Let $X$ be the maximum over all bins of the total weight of the balls
in that bin.  Then for any fixed $c$,
there exists a constant $k$ such that
$X \le k W \lg m / \lg \lg m$ with probability
$1-o(m^{-c})$.
\end{lemma}
\newcommand{\LemmaWeightedBallsInBinsProof}{
The proof is a straightforward generalization of the usual argument
using characteristic functions for for $n=m$ identical balls.

Let $p = 1/b$, $q = 1-p$, and $\alpha > 0$.
Let $X_i$ be a random variable representing
the weight that the $i$-th ball contributes to
some particular bin.
Let $S = \sum_i X_i$.  Then

\begin{eqnarray*}
\E\left[e^{\alpha S}\right]
&=&
\E\left[\prod_i e^{\alpha X_i}\right]
\\
&=&
\prod_i \left(q e^0 + p e^{\alpha w_i}\right)
\\
&\le&
\prod_i \left(1 + p e^{\alpha w_i}\right)
\\
&\le&
\prod_i \exp\left(p e^{\alpha w_i}\right)
\\
&=&
\exp\left(p \sum_i e^{\alpha w_i}\right).
\end{eqnarray*}

Now apply Markov's inequality to get

\begin{eqnarray*}
\Pr\left[S > t\right]
&=&
\Pr\left[e^{\alpha S} > e^{\alpha t}\right]
\\
&<&
\exp\left(p \sum_i e^{\alpha w_i}\right) - \exp(\alpha t)
\\
&=&
\exp\left(p \sum_i e^{\alpha w_i} - \alpha t\right).
\end{eqnarray*}

We now choose the $w_i$ to maximize this quantity subject to the
given constraints $w_i \le W$ and $\sum_i w_i \le bW$.  Observe that
this is equivalent to maximizing $\sum_i e^{\alpha w_i}$.  Since
$e^{\alpha w_i}$ is convex, this is maximized subject to the sum
constraint by setting $w_i = W$ for $i = 1 \ldots b$ and $w_i = 0$
elsewhere.  We thus have

\begin{eqnarray*}
\Pr\left[S > t\right]
&<&
\exp\left(p \sum_i e^{\alpha w_i} - \alpha t\right)
\\
&\le&
\exp\left(p b e^{\alpha W} - \alpha t\right)
\\
&=&
\exp\left(e^{\alpha W} - \alpha t\right).
\end{eqnarray*}

It is not hard to show that the best choice for $\alpha$ is
$\ln(t/W)/W$, giving

\begin{eqnarray*}
\Pr\left[S > t\right]
&<&
\exp\left(e^{\ln(t/W)} - (t/W) \ln(t/W)\right)
\\
&=&
\exp\left((t/W) - (t/W) \ln(t/W)\right).
\\
&=&
\exp\left((t/W)(1 - \ln(t/W)\right).
\end{eqnarray*}

Finally, set $t/W = k \ln m / \ln \ln m$ to get

\begin{eqnarray*}
\Pr\left[S > t\right]
&<&
\exp\left(\frac{k \ln m}{\ln \ln m} \left(1 - \ln k - \ln \ln m + \ln \ln \ln m\right)\right)
\\
&=&
m^{\left\{k \left(\frac{1-\ln k}{\ln\ln m} - 1 + \frac{\ln \ln \ln
m}{\ln \ln m}\right)\right\}}
\\
&=&
m^{-k(1-o(1))}.
\end{eqnarray*}

So the probability that any of the $b$ bins exceeds
$k W \ln m / \ln \ln m$ is at most $bm^{-k(1-o(1))} \le m^{1-k(1-o(1))}$, which is
$o\left(m^{-c}\right)$ for sufficiently large $k$.
}
\ifabs{}{\begin{proof}\LemmaWeightedBallsInBinsProof\end{proof}}

\ifabs{The proof, a straightforward application of characteristic
functions, is given in Appendix~\ref{appendix-balls-in-bins-proof}.}{}
Applying Lemma~\ref{lemma-weighted-balls-in-bins} to the
random-preference algorithm yields:

\begin{theorem}
\label{theorem-random-preference}
The random-preference algorithm achieves a makespan of $O(\log m/\log
\log m)\cdot\OPT$ with high probability for any fixed set $S$ of
available machines.
\end{theorem}
\begin{proof}
Apply Lemma~\ref{lemma-weighted-balls-in-bins} with $b=|S|$ and $W=\OPT$.
\end{proof}

Note that the probability that
Theorem~\ref{theorem-random-preference} fails is polynomial in $m$,
while the number of possible subsets of available machines is
exponential.  So it is possible that the makespan for some particular
subset $S$ is much worse.  This is not a problem if we assume an
oblivious adversary, but may become one if a more powerful adversary
can choose $S$ after determining the algorithm's preference
lists---which it can do by observing the algorithm's behavior.

\section{Algorithm based on binning and hashing}
\label{algorithm-based-on-binning-and-hashing}
\label{section-binhash}

In this section, we introduce an algorithm called \binhash which
achieves constant approximation ratios for both makespan and
reassignment costs.  Unlike the random-preference algorithm of 
Section~\ref{section-makespan-for-random-preferences}, 
the makespan bound is deterministic and holds in all states.  (The
reassignment cost bound is still probabilistic.)

The algorithm is based on the observation that if the number of jobs
$n \le \alpha |S|$ for some constant load factor $\alpha$, we can get
low reassignment costs and (trivially) optimal makespan by assigning each job to
the first empty machine on its preference list.  This process---the
\concept{hashing} step---is
structurally equivalent to hashing with open addressing, and the
number of reassignments caused by adding or removing a job is
bounded by the length of chains in the corresponding hashing
algorithm, which is a constant for fixed $\alpha$.

However, since we cannot assume $n \le\alpha |S|$ in general, we add an
initial \concept{binning} step where jobs are assigned to $\max(1,
\floor{\alpha |S|})$ bins, which are then hashed to machines.  The
binning step sorts jobs by size, and then assigns each job to the
bin whose index, expressed in binary, is the longest available suffix
of the index of the job.  This is a form of round-robin assignment
that, by spreading the jobs roughly uniformly among the bins,
guarantees a constant-factor approximation of the optimal makespan.
At the same time, the number of jobs that move when the number of
bins changes is small, since in addition to guaranteeing an even
spread by size the binning procedure also guarantees an even spread
by job count, and adding or deleting a bin only splits some
existing bin or combines two previous bins.

We now give a formal definition of the algorithm. Given $n$ jobs
with sizes $p_0\ge p_1\ge\cdots\ge p_{n-1}$, and a set $S\in
2^{[m]}/\{\emptyset\}$ of available machines, \binhash computes an
assignment of $n$ jobs to the machines in $S$ in two stages:

\begin{enumerate}
\item Binning stage: Let $b=\max\{\lfloor\alpha|S|\rfloor,1\}$, where
  $\alpha\in(0,1)$ is the load factor parameter of the algorithm.
  Assign the $n$ jobs to $b$ bins by calling
  the function $\mathcal{B}(\{p_0,p_1,\ldots,p_{n-1}\},b)\mapsto\langle
  B^{(b)}_0,B^{(b)}_1,\ldots,B^{(b)}_{b-1}\rangle$, where $B^{(b)}_i$
  is computed by
\begin{eqnarray*}
A_{i} & \leftarrow & \{j=i+k\cdot 2^{\lceil\log{(i+1)}\rceil}\mid
k\ge 0\wedge 0\le j\le n-1\}\\
B^{(b)}_i    & \leftarrow & A_i-\bigcup_{l=i+1}^{b-1}B^{(b)}_l
\end{eqnarray*}
for $i=b-1,b-2,\ldots,0$.

In other words, each bin $B^{(b)}_i$ gets precisely those jobs whose
binary expansions include the binary expansion of $i$ as a suffix,
provided there are no higher-numbered bins that capture them first.
It is immediate from the definition of $B^{(b)}_i$ that every jobs is
assigned to exactly one bin.

\item Hashing stage: The bins are now assigned to the machines in $S$
  in order by a \emph{uniform hashing} function $\mathcal{H}(i,S)$, where
  for each $i$ from $0$ to $\floor{\alpha m}$, bin $i$ is hashed to the
  first machine in a fixed random permutation of all $m$ machines that is
  in $S$ and not occupied by a lower-numbered bin.
\end{enumerate}

We now show that the \binhash algorithm achieves low makespan.  This
follows from the even distribution of the sizes of the bins and the
fact that at most one bin is assigned to each machine.

\begin{lemma}
\label{lemma-binhash-makespan}
Assuming that the optimal makespan of assigning $n$ jobs to $|S|$ machines
is OPT, then
\[
\max_{0\le i\le b-1}|B_{i}^{(b)}|\le \frac{4n}{\alpha|S|}
\]
and
\[
\max_{0\le i\le b-1}{\sum_{j\in B_i^{(b)}}p_j}\le
(1+2/\alpha)\cdot\OPT
\]
for $b=\max\{\lfloor\alpha|S|\rfloor,1\}$.
\end{lemma}
\newcommand{\LemmaBinhashMakespanProof}{
By noting that $B_i^{(b)}$ contains all such $j$ whose longest
available suffix in binary expansion is $i$, it is easy to verify
that,
\begin{eqnarray}\label{eq-bin-size}
B_i^{(b)}\subseteq\{j=i+k\cdot 2^{\lfloor\log{b}\rfloor}\mid k\ge
0\wedge 0\le j\le n-1\}.
\end{eqnarray}

Therefore, for any $0\le i\le b-1$
\begin{eqnarray*}
  |B_{i}^{(b)}| \le  \lfloor\frac{n}{2^{\lfloor\log{b}\rfloor}}\rfloor \le \frac{2n}{b}\le
  \frac{4n}{\alpha|S|}.
\end{eqnarray*}
As for the loads of bins, note that
$OPT\ge\max\{p_0,\frac{1}{|S|}\sum_{j=0}^{n-1}p_j\}$. According to
(\ref{eq-bin-size}) and the non-increasing order of $p_j$, we have
that, for each $0\le i\le b-1$,

\begin{eqnarray*}
  \sum_{j\in B_i^{(b)}}p_j
  & \le & \sum_{k\ge 0}p_{i+k\cdot 2^{\lfloor\log{b}\rfloor}}\\
  & = & p_i+\sum_{k\ge 1}\frac{1}{2^{\lfloor\log{b}\rfloor}}\cdot 2^{\lfloor\log{b}\rfloor}\cdot p_{i+k\cdot 2^{\lfloor\log{b}\rfloor}}\\
  & \le & p_0+\sum_{k\ge 1}\frac{1}{2^{\lfloor\log{b}\rfloor}}\sum_{t=0}^{-1+2^{\lfloor\log{b}\rfloor}}p_{i+k\cdot
  2^{\lfloor\log{b}\rfloor}-t}\\
  & \le & p_0+\frac{1}{2^{\lfloor\log{b}\rfloor}}\sum_{j=0}^{n-1}p_j\\
  & \le & p_0+\frac{2}{b}\sum_{j=0}^{n-1}p_j\\
  & \le & p_0+\frac{2}{\alpha|S|}\sum_{j=0}^{n-1}p_j\\
  & \le & (1+2/\alpha)\cdot OPT.
\end{eqnarray*}
}
\ifabs{

The proof is given in
Appendix~\ref{appendix-lemma-binhash-makespan-proof}.
}{\begin{proof}
\LemmaBinhashMakespanProof
\end{proof}
}

To show low reassignment costs, we must take into account both
reassignments caused by moving bins (when a machine leaves or becomes
available) and reassignments caused by moving jobs between bins.  The
former is bounded by the fact that the number of jobs in each bin is
roughly equal.  The latter requires an analysis of the hashing step.
To simplify the argument, we consider first only the case where a
single machine becomes unavailable.  The case of a new machine
becoming available has the same cost by symmetry, and we will show
later in Theorem~\ref{theorem-binhash-upper-bound}
that the cost of larger transformations can be expressed in
terms of these two cases.

\begin{lemma}
\label{lemma-binhash-reassignment-costs}
For $T\subset S$, and $|T|=|S|-1$, the algorithm \binhash reassigns at
most an expected $\frac{4(2-\alpha)n}{\alpha(1-\alpha)|S|}$ jobs when
moving from state $S$ to state $T$.
\end{lemma}
\begin{proof}
The total reassignments can be upper bounded by the sum of
reassignments caused by binning and the reassignments caused by bin
displacements due to hashing.

According to the definition of binning stage in \binhash, it is easy
to verify  that for any $b\ge 1$,
\begin{eqnarray*}
B^{(b)}_{b-2^{\lfloor\log{b}\rfloor}} &=& B^{(b+1)}_{b-2^{\lfloor\log{b}\rfloor}}\cup
B^{(b+1)}_b\\
B^{(b)}_i &=& B^{(b+1)}_i\mbox{\qquad for }i\neq b-2^{\lfloor\log{b}\rfloor}.
\end{eqnarray*}
Thus the reduction of the last bin is the only case for the
reassignments caused by binning process.

For the single machine in $S\backslash T$, there might be a bin $i$
assigned to it in state $S$. When the state changes from $S$ to $T$, bin
$i$ will be assigned to some other machine in $T$ by the 
hashing step.
This may lead to a recursive displacement of further bins with
larger index than $i$.
Since the placement of bin $i$ in $T$ is uncorrelated with the
preferences of later bins,
the number of such displacements 
can be bounded by the maximum number of bin
displacements caused by inserting an additional bin with an
random index into state $T$'s assignment.

Suppose that $n<m$, and let $\Delta(n,m,i)$ denote the expected
displacements led by inserting an object with index $i\ge 0$ to a
hash table with $m$ slots and $n$ objects by uniform hashing. It is
obvious that all objects with priority $h<i$  will not move,
therefore it is equivalent to assume that all such objects and their
slots are not available to our analysis, thus with probability
$\frac{n-i}{m-i}$ object $i$ hits a slot which is occupied by object
$i'$ where $i'>i$, and then triggers an expected $\Delta(n,m,i')$
displacements led by inserting object $i'$.
\[
\Delta(n,m,i)=\left\{
\begin{array}{lc}
  1 & i=n\\
  1+\frac{n-i}{m-i}\Delta(n,m,i')\\
  \mbox{\quad\qquad for some }i'>i & \mbox{ \emph{o.w.}}.
\end{array}
\right.
\]
By induction, it is easy to show that $\Delta(n,m,i)\le 1/(1-\frac{n}{m})$
for all $i\ge 0$.

Applying this to the bin assignment, we have at most $\Delta(b-1,|S|-1,i)$
displacements of bins, where $b=\max\{\lfloor\alpha|S|\rfloor,1\}$,
therefore,
\[
\Delta(b-1,|S|-1,i)\le\frac{1}{1-\alpha}.
\]
The total reassignment costs in terms of jobs is thus bounded from
above
by $(1+1/(1-\alpha))\max_{0\le i\le
b-1}|B_{i}^{(b)}|\le\frac{4(2-\alpha)n}{\alpha(1-\alpha)|S|}$.
\end{proof}

We now combine the results in Lemmas~\ref{lemma-binhash-makespan}
and~\ref{lemma-binhash-reassignment-costs} to obtain the full result:

\begin{theorem}
\label{theorem-binhash-upper-bound} The following claims hold for any
constant $0<\alpha<1$:

For any $n>0$ and state $S$, assuming that the optimal makespan of
assigning $n$ jobs to $|S|$ machines is OPT, then the makespan of
the assignment obtained by running \binhash with the $n$ jobs on $S$
is within $(1+2/\alpha)\cdot OPT$.

For any states $S$ and $T$, the expected number of reassignments
performed by \binhash going from $S$ to $T$,
\[
\E[\reassignment_B(S,T)] \le
2\left(1+\frac{4(2-\alpha)}{\alpha(1-\alpha)}\right)\cdot
\reassignment^*(S,T),
\]
where the expectation is taken over the randomness of uniform
hashing, and $\reassignment^*$ is defined as in
Theorem~\ref{theorem-normalized-reassignment}.
\end{theorem}
\begin{proof}
Since only one bin is assigned to each machine, the upper bound on
makespan follows directly from Lemma
\ref{lemma-binhash-makespan}.

For the state transition from $S$ to $T$, where $T\subset S$, according to
Lemma \ref{lemma-binhash-reassignment-costs}, we have that
\begin{eqnarray*}
\E[\reassignment_B(S,T)] & \le & \min\left\{n, \frac{4(2-\alpha)n}{\alpha(1-\alpha)}\left(\frac{1}{|S|}+\frac{1}{|S|-1}+\ldots+\frac{1}{|T|+1}\right)\right\}\\
         & \le &
         \min\left\{n, \frac{4(2-\alpha)n}{\alpha(1-\alpha)}\ln\frac{|S|}{|T|}\right\}\\
         & \le &
         \left(1+\frac{4(2-\alpha)}{\alpha(1-\alpha)}\right)(1-|T|/|S|)n\\
         & =& \left(1+\frac{4(2-\alpha)}{\alpha(1-\alpha)}\right)\cdot
         \reassignment^*(S,T).
\end{eqnarray*}
For a general $S$ and $T$, we have that
\begin{eqnarray*}
\E[\reassignment_B(S,T)] & \le & \E[\reassignment_B(S,S\cap T)]+\E[\reassignment_B(T,s\cap T)]\\
         & \le &
         \left(1+\frac{4(2-\alpha)}{\alpha(1-\alpha)}\right)(\reassignment^*(S,S\cap T)+\reassignment^*(T,S\cap T))
         \\
         & \le & 2\left(1+\frac{4(2-\alpha)}{\alpha(1-\alpha)}\right)\cdot
         \reassignment^*(S,T).
\end{eqnarray*}
\end{proof}

It is not hard to show that the coefficient on reassignment costs is
minimized at $\alpha = 2 - \sqrt{2} \approx 0.59\ldots$.  Here the
reassignment costs are bounded by
Theorem~\ref{theorem-normalized-reassignment} at approximately $48.6
\cdot \reassignment^*$ and the makespan at $(3+\sqrt{2})\cdot\OPT
\approx 4.142 \cdot \OPT$.  There are a number of loose
inequalities in the proof of
Theorem~\ref{theorem-normalized-reassignment}, and we believe that a
more careful analysis would show that the
correct minimum coefficient on reassignment costs is closer to $12$ in most
cases.

It is also worth noting that the makespan coefficient $1+2/\alpha$ can
be reduced somewhat by increasing $\alpha$.  However, this dramatically
increases the reassignment costs as $\alpha$ approaches $1$, and the
makespan bound never drops below $3 \cdot \OPT$, which is not
much better than the bound for $\alpha = 2-\sqrt{2}$.
However, it is not out of the question that a more sophisticated
algorithm could achieve higher utilization of the available machines
without blowing up the reassignment costs.

\section{Conclusions and future work}
\label{section-conclusions}

We have described a new problem of path-independent load balancing for
unreliable machines, where the goal is to minimize makespan while
simultaneously minimizing the cost of reassigning jobs from one
machine to another subject to the constraint that assignments cannot
depend on the previous history.  We have also obtained some initial
results showing that it is possible to achieve constant approximation
ratios to both the optimal makespan and optimal reassignment costs.

However, much work still needs to be done.  The proven constant approximation
ratios for the \binhash algorithm---particularly for makespan---are
still quite high, and it would be useful to have an algorithm with
better constants.

The assumption of identical machines is a strong
one. It is not clear whether our results can be generalized to the
case of uniform machines (where different machines have different
capacities) or to the even more general case of nonuniform machines
(where different jobs may have different effective sizes on different
machines).  This last case may be particularly important in
interdomain routing, as particular flows may be forbidden from traveling
over certain pipes by contractual requirements or security concerns.

Finally, we have made very generous assumptions
about the nature of the jobs and the nature of the adversary.  It
would be interesting to determine whether it is possible to solve
path-independent load balancing with jobs that vary over time or with
a more powerful adversary that can observe and respond to the
algorithm's actions.

\typeout{end of main body}

\bibliographystyle{abbrv}
\bibliography{paper}

\begin{abs}
\appendix

\section{Proof of Theorem~\ref{theorem-randomized-lower-bound}}
\label{appendix-randomized-lower-bound-proof}

\TheoremRandomizedLowerBoundProof

\section{Proof of Lemma~\ref{lemma-weighted-balls-in-bins}}
\label{appendix-balls-in-bins-proof}

\LemmaWeightedBallsInBinsProof

\section{Proof of Lemma~\ref{lemma-binhash-makespan}}
\label{appendix-lemma-binhash-makespan-proof}

\LemmaBinhashMakespanProof

\end{abs}

\end{document}